\documentclass[11pt]{article}
\usepackage{graphicx}
\usepackage{wrapfig}
\usepackage{subcaption}
\usepackage{amssymb}
\usepackage{amsmath}
\usepackage{color}
\usepackage{caption}

\begin{document}

\baselineskip=24pt

\begin{titlepage} 

\centerline{\bf Energy dynamics and current sheet structure in fluid and kinetic simulations}
 \centerline{\bf of decaying magnetohydrodynamic turbulence} 

\bigskip

\centerline{K. D. Makwana$^{1,\ast}$, V. Zhdankin$^2$, H. Li$^3$, W. Daughton$^3$, F. Cattaneo$^1$}
\centerline{$^1$\emph{Department of Astronomy and Astrophysics, University of Chicago, Chicago, IL-60637}}
\centerline{$^2$\emph{Department of Physics, University of Wisconsin-Madison, Madison, WI-53706}}
\centerline{$^3$\emph{Los Alamos National Laboratory, Los Alamos, NM-87544}}

\bigskip


\medskip

\begin{abstract}
Simulations of decaying magnetohydrodynamic (MHD) turbulence are performed with a fluid and a kinetic code. The initial condition is an ensemble of long-wavelength, counter-propagating, shear-Alfv\'{e}n waves, which interact and rapidly generate strong MHD turbulence. The total energy is conserved and the rate of turbulent energy decay is very similar in both codes, although the fluid code has numerical dissipation whereas the kinetic code has kinetic dissipation. The inertial range power spectrum index is similar in both the codes. The fluid code shows a perpendicular wavenumber spectral slope of $k_{\perp}^{-1.3}$. The kinetic code shows a spectral slope of $k_{\perp}^{-1.5}$ for smaller simulation domain, and $k_{\perp}^{-1.3}$ for larger domain. We estimate that collisionless damping mechanisms in the kinetic code can account for the dissipation of the observed nonlinear energy cascade. Current sheets are geometrically characterized. Their lengths and widths are in good agreement between the two codes. The length scales linearly with the driving scale of the turbulence. In the fluid code, their thickness is determined by the grid resolution as there is no explicit diffusivity. In the kinetic code, their thickness is very close to the skin-depth, irrespective of the grid resolution. This work shows that kinetic codes can reproduce the MHD inertial range dynamics at large scales, while at the same time capturing important kinetic physics at small scales. 

\vskip10ex

\noindent $^{\ast}$kirit.makwana@gmx.com

\baselineskip=24pt

\end{abstract}

\end{titlepage}

\setcounter{page}{2}

\noindent{\bf I. Introduction}

Magnetohydrodynamic (MHD) turbulence describes the phenomenon of turbulence in conducting fluids~\cite{biskamp2003}. It is observed in a variety of systems, ranging from laboratory plasmas~\cite{moreau1998} to space and astrophysical plasmas~\cite{zhoumatthaeus2004}. Similar to hydrodynamic turbulence~\cite{kolmogorov1941}, three dimensional MHD turbulence is characterized by an inertial range where energy is forward cascaded from larger to smaller scales conservatively~\cite{kraichnan1965}. The inertial range spectrum of incompressible, 3-dimensional MHD turbulence is well-studied and several theories have been proposed to explain it. Goldreich and Sridhar~\cite{goldreichsridhar1995} introduced the concept of critical balance between linear wave period and nonlinear interaction timescale, leading to an anisotropic cascade of energy that is faster in the perpendicular scales compared to parallel scales. 

As energy cascades to smaller scales, at small enough scales it has to be dissipated into heat. Dissipation of the MHD turbulence cascade is an important source of heat in space and astrophysical plasmas. It has been cited to explain heating of the solar corona and solar wind~\cite{leamonmatthaeus2000}, as well as a possible source of heat in the intracluster medium to balance radiative cooling~\cite{zhuravlevachurazov2014}. However, dissipation of MHD turbulence is still not well-understood. In hydrodynamics, dissipation occurs at small length scales due to viscosity which can be understood from molecular theory. However, many astrophysical plasmas are weakly collisional at small scales. Collisionless damping mechanisms resulting from wave-particle interactions have been proposed to explain dissipation at such scales~\cite{howescowley2008}.  MHD is a fluid model for plasmas which is strictly applicable only at length scales larger than the ion gyro-radius, $\rho_i=\sqrt{k_{B}m_iT_i}/(q_iB)$~\cite{schekochihincowley2009}. Here $k_B$ is the Boltzmann constant, $m_i$ is the ion mass, $T_{i}$ is the ion temperature, $q_i$ is the ion charge, and $B$ is the magnetic field strength. Thus, some theories propose that the energy cascade turns into a kinetic-Alfven wave cascade at scales smaller than the ion gyro-radius~\cite{howesdorland2008}. Another possibility would be conversion to a Whistler wave cascade~\cite{shaikhzank2009}. In either case, we expect the energy to be dissipated away into heat at the electron gyro-radius scale.

Another suggested mechanism for dissipation is nonlinear processes in thin current sheets~\cite{dmitrukmatthaeus2004}. In numerical simulations of MHD turbulence with a mean field, one often sees formation of current sheet structures. Boldyrev~\cite{boldyrev2005} has introduced the concept of dynamic alignment, in which the velocity and magnetic fluctuations in the plane perpendicular to the mean field become increasingly aligned at smaller scales, producing sheet-like structures rather than filaments. Such structures can dissipate energy intermittently~\cite{wanmatthaeus2012}. Some resistive MHD simulations have shown strong current sheets that occupy 1\% of the simulation volume but account for 25\% of the resistive dissipation~\cite{zhdankinuzdensky2013}. However, the thickness of these current sheets depends on the resistivity supplied by the user, or on the grid spacing if sufficient resistivity is not provided. In either case, the dissipation calculated in such current sheets is not based on real micro-physics and hence the use of kinetic codes is essential to analyze dissipation in such sheets. 

Recently particle-in-cell and gyro-kinetic codes have started probing the dissipation range of the MHD cascade. Such simulations also show the formation of electron-length-scale current sheets~\cite{karimabadiroytershteyn2013,tenbargeapjl2013,tenbargehowes2013}. However, it has been difficult to produce the entire spectrum of energy cascade in these simulations, starting from the inertial range going up to the dissipative range. In such a case, an antenna can be used to model the injection of energy at the scale of the simulation domain,  mimicking the cascade of energy down from larger scales~\cite{tenbargehowes2014}. 

In this work, we simulate the entire spectrum of MHD turbulence, from inertial to dissipative range, using a particle-in-cell  (PIC) kinetic code. This allows us to describe the energy input from large scales without requiring an artificial setup. We also run the same simulations with a compressible MHD code and compare the results between the two codes at all scales. This gives a check on the validity of PIC codes in the continuum limit at large scales. Comparison at small scales provides us with clues about where kinetic physics plays an important role. We focus on large scale dynamics and current sheets in this study. In Sec. II we describe the setup of the numerical simulations. Energy is injected at large scales by means of strongly interacting shear-Alfv\'{e}n waves at the initial time. The turbulence is then allowed to develop and decay, as there is no continuous forcing. In Sec. III we report the results of global energy dynamics and energy spectra. The turbulent energy decay rates are remarkably similar in both codes, despite the very different dissipation mechanisms present in them. We also find that the spectra match very well in the inertial range, especially when larger simulation domains are utilized in the PIC code. The dissipative range spectra show important differences between the PIC and MHD simulations. In Sec. IV we describe an analysis of the geometrical characteristics of current sheets. They show similar lengths and widths in the two codes which scale with the outer (driving) scale of turbulence. All these observations validate that  the PIC code can reproduce MHD dynamics at large scales. However, at small scales we see critical differences. The thickness of current sheets is determined by the grid resolution in MHD, due to the absence of any explicit diffusivity. However in the PIC code, current sheets always show thickness close to the skin-depth, thus capturing the relevant kinetic physics. We end with discussion in Sec. V.

\bigskip

\noindent{\bf II. Setup of numerical simulations}

We simulate decaying turbulence with the use of two codes - the MHD code PLUTO~\cite{mignonebodo2007}, and the particle-in-cell code VPIC~\cite{bowersalbright2008}. PLUTO solves the ideal MHD equations with a finite-volume technique. We do not specify any explicit diffusivity, thereby relying solely on the numerical diffusion of the finite grid spacing. In the presence of a mean magnetic field $B_0$, we can define an Alfven velocity ($v_{A}=B_0/\sqrt{\mu_0\rho}$). Here $\mu_0$ is the permeability of free space and $\rho$ is the plasma mass density. The ideal MHD equations do not have a natural length scale and can be normalized to an arbitrary length scale $L$, thereby also specifying a time scale. The only parameter that remains in the dimensionless MHD equations is the plasma beta, $\beta=p_{tot}/(B_0^{2}/2\mu_0)$, where $p_{tot}$ is the total pressure. We choose $\beta=0.33$. The adiabatic equation of state is used. 

VPIC integrates the coupled Maxwell-Boltzmann equations~\cite{bowersalbright2008}. We consider a pair plasma for the sake of computational ease. Since VPIC is a particle-in-cell code, we have to specify an initial velocity distribution function. We give the particles a Maxwellian distribution with a thermal speed $v_{th}=\sqrt{k_BT_i/m_i}=\sqrt{k_BT_e/m_e}=0.08c$ ($c$ is the speed of light). There are 350 particles of each species in every cell initially. The plasma beta is matched between the two codes by selecting the ratio of electron plasma frequency, $\omega_{pe}$, to the electron cyclotron frequency, $\omega_{ce}$, as $\omega_{pe}/\omega_{ce}=3.6$. The electron cyclotron frequency is defined as $\omega_{ce}=|q_eB_0/m_e|$ and the electron plasma frequency is $\omega_{pe}=\sqrt{n_0e^2/(\epsilon_0m_e)}$, where $n_0$ is the electron number density and $\epsilon_0$ is the permittivity of free space. The ratios $v_{th}/c$ and $\omega_{pe}/\omega_{ce}$ are the only two parameters relevant in the dimensionless form of the kinetic equations. The above choice of these two parameters gives $\beta=0.33$, ensuring that both codes are simulating the same physical plasma.

In both codes the velocity is normalized by speed of light. In PLUTO, the magnetic field is normalized by $c\sqrt{4\pi\rho_0}$, where $\rho_0$ is the mean density in C.G.S. units. In VPIC the magnetic field is normalized by $\omega_{pe}m_e/e$, where all quantities are in S.I. units. However, for all the energy quantities plotted in this paper, the energies calculated in both codes are converted to S.I. units for direct comparison. In order to do this, the mean magnetic field is given the value $B_0=10G$. This then gives the S.I. value for all the other quantities using the dimensionless parameters specified.

The simulation domain is a three-dimensional, rectangular box with periodic boundary conditions. The box is four times longer along the $z$ direction, parallel to the background magnetic field, with the knowledge that current sheets will be elongated along the background field. The initial condition is specified by an ensemble of shear-Alfven waves superimposed on the plasma. In other words, the initial perturbations of the magnetic and velocity fields are of the form of shear-Alfven wave eigenvectors, as shown below,
\begin{equation}
\delta \mathbf{b} = \sum_{j,l} a_0B_0\cos (jk_zz+lk_yy+\phi_{j,l})\hat{x}+ \sum_{m,n} a_0B_0\cos (mk_zz+nk_xx+\phi_{m,n})\hat{y},
\label{deltab}
\end{equation}
\begin{equation}
\delta \mathbf{v} = -\sum_{j,l} sgn(j)a_0v_A\cos (jk_zz+lk_yy+\phi_{j,l})\hat{x}- \sum_{m,n} sgn(m)a_0v_A\cos (mk_zz+nk_xx+\phi_{m,n})\hat{y}.
\label{deltav}
\end{equation}
Each $(j,l)$ or $(m,n)$ combination specifies a shear-Alfven wave. The wavenumbers take the values $(j,l)=(1,1); (1,2); (-2,3)$ and $(m,n)=(-1,1); (-1,-2); (2,-3)$ with $k_{x,y,z}=2\pi/L_{x,y,z}$. The amplitude of the perturbations is $a_0=0.18$. When the $sgn(j)$ or $sgn(m)$ is negative, then we have a shear Alfven wave moving in positive-$z$ direction, and vice-versa. We see that there are 3 waves moving in positive $z$ direction, and 3 moving in the negative $z$ direction, with equal energy in both directions, implying balanced turbulence. The interaction of these waves quickly generates MHD turbulence. In VPIC, velocity perturbations are incorporated by defining the initial Maxwellian distribution functions to be centered around the space-dependent mean velocity perturbations. Since we are considering a pair plasma, the mean velocity of each species contributes equally to the fluid velocity. This mean velocity is derived from the velocity perturbations in Eq.~\ref{deltav}, plus a component for the current densities needed to produce the perturbed magnetic fields in Eq.~\ref{deltab}. In addition, electric fields have to be provided in VPIC, consistent with the ideal MHD equation $\mathbf{E}=-\mathbf{v}\times\mathbf{B}\approx -\delta\mathbf{v}\times\mathbf{B}_0$ (here we ignore the weaker $\delta \mathbf{v} \times \delta \mathbf{b}$ term). 

The initial r.m.s. amplitude of the perpendicular magnetic field perturbations is $b_{\perp,\mathrm{rms}}/B_0=0.31$. Similarly the initial r.m.s. amplitude of the velocity perturbations is the same for shear-Alfven waves, $v_{\mathrm{rms}}/v_{A}=0.31$, which is the initial r.m.s. Alfvenic Mach number ($M_A$). The initial r.m.s. sound Mach number is $M=v_{\mathrm{rms}}/c_s=0.59$. The perpendicular wavenumber of the Alfven-wave ensemble is roughly 4 times larger than the parallel wavenumbers, due to the box aspect ratio of 1:4. This satisfies the condition of critical balance, $k_{\perp}v_{\perp}\sim k_{\parallel}v_A$, at the outer scales which helps in quickly setting up a strong turbulence cascade.  

Different simulations are carried out keeping the same initial conditions, but changing the box size and resolution. We define the resolution as the number of cells per unit length, so the resolution in $x$-direction would be $N_x/L_x$, where $N_x$ is the number of cells in $x$-direction and $L_x$ is the corresponding box length. Table~\ref{runs} shows the settings for the different runs performed. We take the same number of cells ($N_x=N_y=N_z$) in all the three directions. This is possible because of the nature of the anisotropic cascade, which transfers energy to smaller scales preferentially in the perpendicular direction, thus allowing us to get away with $1/4^{th}$ the resolution in $z$. We also find that the current sheets are elongated along the $z$ direction and do not require as much parallel resolution as in the perpendicular directions. We have verified that the results do not change with higher resolution in $z$. Run I serves as the base case. Run 0 is a test case with the lowest resolution. Run II has double the resolution of run I. Run III has same resolution as run I, but double the box size. Run IV has the same resolution as run 0, but the box is 4 times larger than run I. The box length and resolutions have been chosen such that there are at least two decades between the initially-driven scales and kinetic scales for an inertial range to form, but the kinetic scales are also resolved at the same time. Changing resolution and box sizes will tell us about how the outer and inner scales affect turbulence properties. It should be pointed out that runs II, III, and IV are physically the same in ideal MHD, because changing box size just involves changing the length normalization factor in ideal MHD. However, in the kinetic formulation length has a physical meaning and hence runs II, III, and IV are physically different runs in VPIC. All the simulations are run up to 3 Alfven crossing times ($\tau_A=L_z/v_A$), with data being output every $\Delta t=0.2\tau_A$. We first look at the energetics and turbulent spectra of these simulations.

\begin{table}
\begin{center}
\begin{tabular}{|c|c|c|c|c|}
\hline
Run & $L_x(d_i)$ & $L_y(d_i)$ & $L_z(d_i)$ & $N_{x,y,z}$ \\
\hline
0 & 120 & 120 & 480 & 288 \\
\hline
I & 120 & 120 & 480 & 576 \\
\hline
II & 120 & 120 & 480 & 1152 \\
\hline
III & 240 & 240 & 960 & 1152 \\
\hline
IV & 480 & 480 & 1920 & 1152\\
\hline
\end{tabular}
\caption{Settings for simulations. The box lengths are specified in terms of ion-skin-depth, $d_i$. $N_{x,y,z}$ specifies the number of cells in each direction. There are equal number of cells in each direction.}
\label{runs}
\end{center}
\end{table}

\bigskip

\noindent{\bf III. Energy dynamics and turbulent spectra}

\begin{figure}[h]
\centering
\resizebox{0.65\textwidth}{!}{
\includegraphics{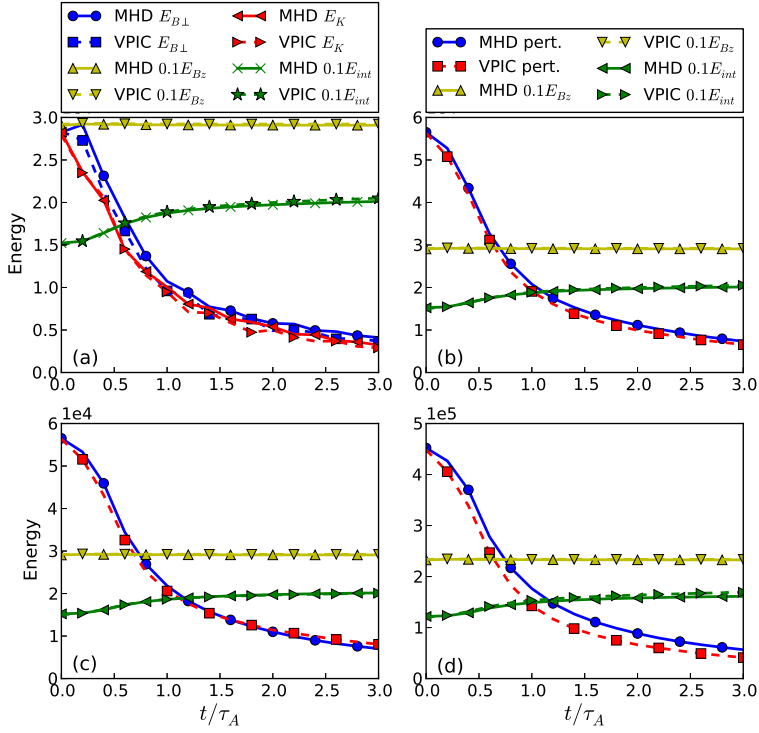}}
\caption{(color online) Energy evolution for simulation runs I, II, and III. (a) shows the perpendicular magnetic field energy (in blue) and kinetic energy (in red), internal energy (in green), and parallel magnetic field energy (yellow) for run I. The solid lines are from PLUTO simulations and dashed lines are from VPIC. The energy axis is normalized by $10^{4}$, which also applies to figures (b) and (c). The internal energy and the energy in parallel magnetic field is multiplied by $0.1$ for ease of viewing. (b) shows the total perturbation energy for PLUTO (in blue) and VPIC (in red) for run I. Internal and background magnetic field energies are shown as in (a). The legend for (b) is followed in (c) and (d). (c) shows energy terms in run II. (d) shows energy terms for run III, its energy axis is normalized by $10^5$.}
\label{paper_energy}
\end{figure}

Let us first look at the energetics of the system. Fig.~\ref{paper_energy} plots the various energy components of the plasma in simulation runs I, II, and III, of both the codes. In the initial state, magnetic and kinetic perturbation energies are equal, because we have only shear Alfven waves, and add up to 19.4\% of the background magnetic field energy. As time advances, the shear-Alfven waves interact, produce turbulence, which forward cascades the perturbation energy to small scales, and ultimately dissipates into heat. We see this clearly in Fig.~\ref{paper_energy}. The total energy is conserved in PLUTO code to better than 0.5\% error, and in VPIC energy conservation is better than 3\% error for these runs. Thus, the internal energy correspondingly increases as the magnetic and kinetic energy decay. PIC codes have a finite-grid instability which results in numerical heating of the plasma leading to an increase in internal energy. Typically this instability kicks in when the cell size ($\Delta x$) becomes larger than $3\lambda_{De}$, where $\lambda_{De}$ is the electron Debye length. This is true for all our runs in the parallel direction and for runs 0 and IV in the perpendicular direction. In the appendix we study how much heating we are getting due to this instability and find that it is very weak and should not change our results in any significant way. 

The internal energy in PLUTO is given by $\int d^3x[p/(\gamma-1)]$, with $\gamma=5/3$. The internal energy in VPIC is defined as the total kinetic energy of all particles minus the kinetic energy calculated from the fluid velocity obtained by averaging the particle velocities in each cell. These two internal energies match very nicely as seen in Fig.~\ref{paper_energy}. The energy in the parallel field along $z$ direction remains constant. This energy includes, for the most part, the mean field $B_0$, plus a tiny fraction (0.3\% of total energy) from parallel magnetic fluctuations which develop with the turbulence. In Fig.~\ref{paper_energy}(a) we see that the perpendicular magnetic field energy in MHD behaves differently compared to VPIC during the first step from $t/\tau_A=0$ to $t/\tau_A=0.2$. We see a slight increase in the magnetic energy during this time for MHD. This behavior is seen in the VPIC simulation also, but for a time shorter than $\Delta t/\tau_A=0.2$ before the magnetic energy starts decreasing, so it is not visible in the figure. This might happen due to small-scale magnetic fields being wound up by the velocity perturbations as the turbulence develops. These fluctuations could be damped faster in VPIC due to kinetic damping, whereas they would be undamped in MHD. Below we will see indications of this occurring. Correspondingly, the internal energy in the MHD code increases slightly slower than VPIC during this time. 
After $t/\tau_A=0.2$, both codes show remarkably similar behavior, with almost the same energy decay rates in all the three simulation runs shown in Figs.~\ref{paper_energy}(b), (c), and (d). Overall the differences between the two codes are less than 1\% of the total energy, which is within the limits of error in energy conservation for these simulations, and hence the decay rates are the same within error bars. If we fit a power law to the energy decay plot, we find a decay rate very close to $E\sim t^{-1}$, where $E$ is the sum of turbulent magnetic and kinetic energy.

The decay rates are the same despite the fact that dissipations mechanisms are very different in the two codes. We depend solely on numerical diffusivity for dissipation in MHD, whereas the kinetic code has access to all kinetic damping mechanisms. One possible reason for this could be the invariance of energy transfer rate across the inertial range of turbulence. In this picture, the rate of energy transfer across each scale in the inertial range is determined by the nonlinearity and is an invariant. There is a possibility of non-local interactions in MHD turbulence where energy is transferred directly from large to small scales without passing through each intermediate scale~\cite{alexakismininni2005}. However, this non-local coupling should weaken as the dynamical range of our simulations increases~\cite{aluieeyink2010}, giving similar local cascade rates in both the codes. This energy is dissipated at the small scales at the same rate, with the small scale structures adjusting themselves to provide the needed rate of dissipation. As a result, the small scale physics does not impact the energy decay rate. To check this we next look at the energy spectrum of this turbulence. 

\begin{figure}[h]
\centering
\resizebox{0.5\textwidth}{!}{
\includegraphics{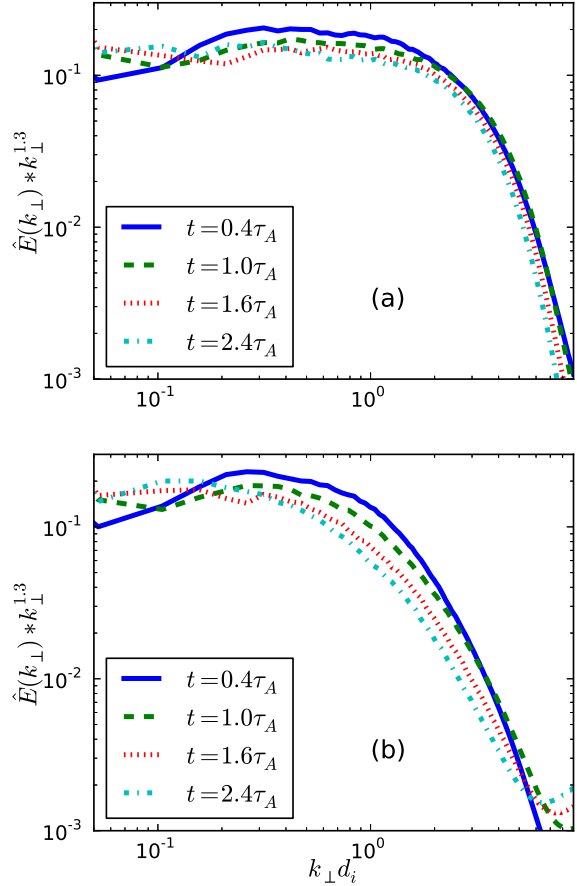}}
\caption{(color online) The total energy spectrum normalized by the total fluctuation energy at different time steps in run II. (a) is for the PLUTO run and (b) is for the VPIC run. The spectrum is also multiplied by $k_{\perp}^{1.3}$.}
\label{sp_time_normed}
\end{figure}

 Fig.~\ref{sp_time_normed} shows the perpendicular wavenumber power spectra for the total energy at different times in run II for both codes. The total energy spectrum is defined as $(|\hat{\mathbf{b}}(\mathbf{k}_{\perp})|^2+|\hat{\mathbf{v}}(\mathbf{k}_{\perp})|^2)/2$, where $\hat{\mathbf{b}}(\mathbf{k}_{\perp})$ is the Fourier transform of the magnetic field $\mathbf{B}$ (in normalized units) in the perpendicular direction to the mean field, averaged over the parallel direction. Similarly $\hat{\mathbf{v}}(\mathbf{k}_{\perp})$ is the Fourier transform in perpendicular direction, averaged over parallel direction, of the field $\sqrt{\rho}\mathbf{v}$ (in normalized units), with the $\sqrt{\rho}$ ($\rho$ is the density) factor coming in due to the compressible nature of our simulations. The fluid velocity $\mathbf{v}$ is a dependent variable in MHD, whereas in VPIC it is derived by averaging the particle velocities in each cell.  
As this is a decaying turbulence, the spectrum changes as a function of time, with the total energy decreasing monotonically. To overcome this, in Fig.~\ref{sp_time_normed} we have normalized the spectrum by the total fluctuation energy at each time step, thus it shows the same energy level at each time step. Initially, the smallest 3 perpendicular wave-numbers are populated (as specified in Eqs.~\ref{deltab}-\ref{deltav}) and the higher wave numbers have no energy. As the simulation proceeds, higher wave numbers get populated rapidly (within $t=0.2\tau_A$). By $t=0.4\tau_A$, the spectra appear to develop a power law behavior in both codes. However, there are some important differences between the two codes. As we go further in time, the shape of the spectrum, including its power law slope, remains very similar in Fig.~\ref{sp_time_normed}(a) which is for MHD. In fact the inertial range gets extended into low-k wavenumbers as time goes on. For MHD, the inertial range behavior extends from $k_{\perp}d_i\approx 0.2$ to $k_{\perp}d_i\approx 1.0$ and we see a slope of $k_{\perp}^{-1.3}$. On the other hand for VPIC in Fig.~\ref{sp_time_normed}(b), the extent of inertial range is narrower from $k_{\perp}d_i\approx 0.3$ to $k_{\perp}d_i\approx 0.8$ at $t=0.4\tau_A$. Its slope is also steeper compared to the MHD run. Moreover, the shape of the VPIC spectrum changes significantly as time goes on. The extent of the power law behavior decreases drastically by $t=2.4\tau_A$, and the entire spectrum shifts to the left. This indicates that dissipation is active even at low wavenumbers ($k_{\perp}d_i<1.0$) in VPIC, which is not the case in MHD. This is a sign of kinetic damping mechanisms which are active at all wavenumbers, but their damping rate increases with $k_{\perp}$. Although damping at low $k_{\perp}$ is weak, as time progresses it starts to show an effect, which is what we are presumably seeing in Fig.~\ref{sp_time_normed}(b). Below, we will further investigate whether kinetic damping can explain the dissipation in VPIC.

\begin{figure}[h]
\resizebox{1.0\textwidth}{!}{
\includegraphics{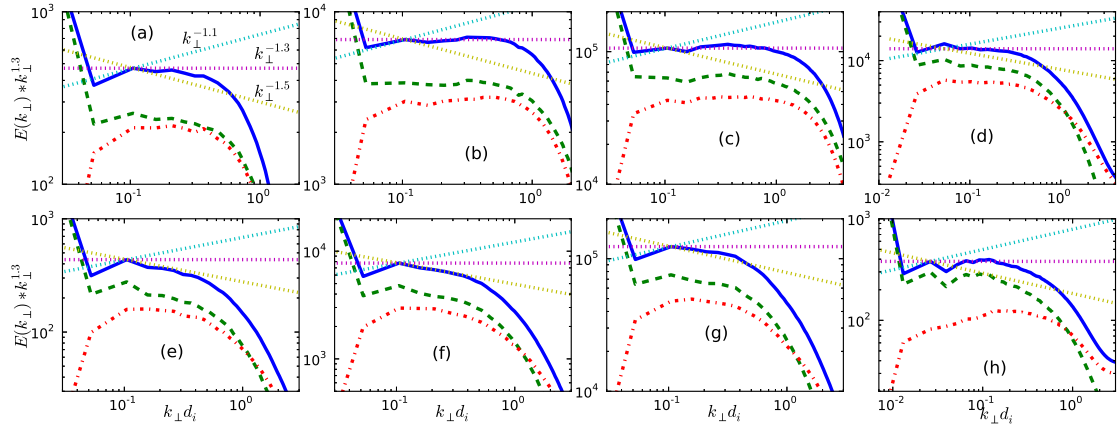}}
\caption{(color online) (a) MHD Run 0, (b) MHD Run I, (c) MHD Run II, (d) VPIC Run III, (e) VPIC Run 0, (f) VPIC Run I, (g) VPIC Run II, and (h) VPIC run IV. Solid blue line is the total energy (magnetic+kinetic) spectrum, dashed green line is the magnetic energy spectrum, and red dash-dotted line is the kinetic energy spectrum. All spectra are multiplied by a factor of $k_{\perp}^{1.3}$, with the dotted lines showing the reference slopes of $k_{\perp}^{-1.1}$, $k_{\perp}^{-1.3}$, and $k_{\perp}^{-1.5}$ as labelled in fig (a).}
\label{kperp_all}
\end{figure}

The time evolving spectrum in Fig.~\ref{sp_time_normed} is time-averaged by compensating for the decay in energy at every time step by a multiplying factor which is the ratio of average energy in the range $0.2\le k_{\perp}d_i \le 0.8$ ($0.1\le k_{\perp}d_i \le 0.4$ for run IV) at $t=1.0\tau_A$ to the average energy in the same range at the given time step. The averaging is performed over 11 time-slices from $t=1\tau_A$ to $t=3\tau_A$. This averaged spectrum is shown in Fig.~\ref{kperp_all} for the different simulations.  
Despite the complication introduced by decaying turbulence and the limited time-averaging, we get convergent power law behavior. Figs.~\ref{kperp_all}(a), (b), and (c) show spectra from MHD runs 0, I, and II, respectively which have resolution of $288^3$, $576^3$, $1152^3$ respectively, with the box size normalized to $(120d_i,120d_i,480d_i)$. We see that the total energy spectrum is well-converged to a slope of $k_{\perp}^{-1.3}$. The magnetic energy is slightly larger than the kinetic energy. The MHD spectra for runs III and IV are not shown as they show very similar spectra to run II. This is because they share the same number of grid cells, only differ in box size. In ideal MHD there is no natural length scale, and hence the box size is just a normalization constant chosen arbitrarily by the user. Runs III and IV are the same computation as run II, except that all length quantities in them get multiplied by factors of 2 and 4 respectively.    

The spectra from VPIC simulations are shown in Figs.~\ref{kperp_all} (e), (f), (g), (d), and (h) for runs 0, I, II, III, and IV respectively. Figs~\ref{kperp_all}(e) and (f) show a steeper total energy spectrum of $k_{\perp}^{-1.5}$. These runs are with lower resolution and the smallest box size. Compared to MHD spectra in Figs.~\ref{kperp_all}(a) and (b) where the power law turns over at around $k_{\perp}d_i\approx 0.6$, here it turns over slightly earlier at $k_{\perp}d_i\approx 0.4$, giving a shorter inertial range. Like in MHD, the magnetic energy is slightly larger than kinetic energy for $k_{\perp}d_i\lesssim1.0$, but this trend reverses for $k_{\perp}d_i\gtrsim 1.0$, unlike in MHD. This indicates a kinetic damping mechanism acting to suppress the magnetic field fluctuations at large wavenumber. This can be the reason for not observing a magnetic energy spike in VPIC at $t/\tau_A=0.2$ in Fig.~\ref{paper_energy}(a). Fig.~\ref{kperp_all}(g) is the highest resolution run with $1152^3$ cells on a box size $(120d_i,120d_i,480d_i)$. It also shows a spectrum close to $k_{\perp}^{-1.5}$, showing that the spectrum is converged for this box size, even at the resolution of run 0. Fig.~\ref{kperp_all}(d) is for run III in which the box lengths are increased by a factor of two to $(240d_i,240d_i,960d_i)$, which unlike in MHD, means a physical change for VPIC. Now we see the total spectrum is much closer to the MHD result of $k_{\perp}^{-1.3}$. For run IV, shown in Fig.~\ref{kperp_all}(h), the box size is again doubled to $(480d_i,480d_i,1920d_i)$, thus giving a resolution equivalent to that of run 0. Here the spectrum is completely converged with the MHD result of $k_{\perp}^{-1.3}$. So we recover the fluid result when a larger box size is used. Conversely, the MHD simulations correctly describe the dynamics of a plasma at large scales. 

This observed slope of $k_{\perp}^{-1.3}$ is shallower compared to strong turbulence results from incompressible MHD turbulence studies~\cite{perezboldyrev2008}. This raises the question whether the numerical simulations are spoilt in some manner. However, the discrepancy could be due to the decaying nature of this turbulence. The slope of $k_{\perp}^{-1.3}$ is close to $k_{\perp}^{-4/3}$. This slope is consistently observed in all our MHD simulations irrespective of the resolution, showing that it is converged. In VPIC we observe a steeper slope of $k_{\perp}^{-3/2}$ in our smaller box (runs 0, I, II), but again it is converged w.r.t. resolution. The slope in VPIC also becomes $k_{\perp}^{-4/3}$ when we take the two larger boxes. The fact that the slope in a PIC simulation, which is numerically very different from a MHD simulation, converges to the MHD value in the large box limit gives us confidence that the MHD result is meaningful. Secondly, other simulations of decaying turbulence have also reported this slope. In Ref.~\cite{salvesenbeckwith2014} a decaying magnetized Kelvin-Helmholtz instability is simulated. Although it is a different setup, the 1-D wavenumber spectrum was observed to be $k^{-4/3}$. Lastly, we have performed numerical simulations of forced MHD turbulence with the same MHD code in the same setup, with only the addition of a forcing term. In that case, the perpendicular wavenumber spectrum comes out to be $k_{\perp}^{-3/2}$, which is in accordance with the result of forced MHD turbulence in Ref.~\cite{perezboldyrev2008}. All these observations give us confidence that the spectral slope result is not due to spoilt numerical simulations. The phenomenological explanation of why one is getting this slope is not understood and should be explored in future work.

We have seen almost the same rate of energy decay in both PIC and MHD codes. A natural explanation for this is that the small scales adjust themselves to dissipate and balance whatever energy is cascaded down from larger scales. This also seems likely because we see very similar inertial range spectra in both codes. But how do the small scales adjust themselves to do the required dissipation? In MHD, the numerical grid is responsible for this dissipation, either due to shocks or numerical reconnection. In VPIC, collisionless damping can play a role. We will investigate detailed dissipation mechanisms in future work, as this work is more concerned with characterizing the larger scale dynamics. However, we can roughly estimate if collisionless damping in VPIC can explain the observed dissipation rate. Ref.~\cite{garykarimabadi2009} calculated linear, collisionless damping rates for pair-plasmas. We use these damping rates to estimate kinetic damping in our VPIC simulations. We separate the total energy into two parts, a low-k part ($E_{<}$) and a high-k part ($E_{>}$), defined as,
\begin{equation}
E_{<}=\int_{0}^{k_0}E(k_{\perp})dk_{\perp},
\end{equation} 
\begin{equation}
E_{>}=\int_{k_0}^{\infty}E(k_{\perp})dk_{\perp}.
\end{equation}
We assume kinetic damping is active at wavenumbers larger than $k_0$. The rate at which energy enters this high-k region is estimated by taking the time derivative of $E_{<}$, assuming that all this energy cascades forward to $k_{\perp}>k_0$. Thus, this quantity is defined as,
\begin{equation}
\frac{dE_{<}}{dt}=\frac{E_{<}(t_2)-E_{<}(t_1)}{t_2-t_1},
\end{equation}
where $t_1$ and $t_2$ are two neighboring time steps. This quantity gives an upper limit on the nonlinear energy cascade rate, because it assumes that all this energy goes into the $k_{\perp}>k_0$ region. This has to be balanced by some dissipation mechanism in the high-k region. This is estimated as follows,
\begin{equation}
\gamma E_{>}=\int_{k_0}^{\infty}\gamma(k_{\perp})[(E(k_{\perp},t_1)+E(k_{\perp},t_2))/2]dk_{\perp},
\end{equation}
where $\gamma(k_{\perp})=\gamma(k_0,k_{\parallel 0})(k_{\perp}/k_0)^{\alpha}$ is a damping rate. Ideally, the accurate way to calculate the energy dissipation by collisionless damping would be to use the exact value of $\gamma(k_{\parallel},k_{\perp})$ obtained from a linear Vlasov solver. This will be a topic of future study, however here we are trying to obtain a crude estimate of collisionless damping, and hence we use this simple function for $\gamma$. All the wavenumber variables are implicitly taken to be normalized with skin-depth ($d_i=d_e$). The constant $\gamma(k_0,k_{\parallel 0})$ is taken from Eq. (2) of Ref.~\cite{garykarimabadi2009}, which is reproduced here,
\begin{equation}
\frac{\gamma(k_0,k_{\parallel 0})}{|\Omega_e|}=-A(\beta_e)k_0^2k_{\parallel 0}.
\end{equation}
This expression is strictly valid only at $|k_0|,|k_{\parallel 0}|\ll 1.0$. However, looking at Fig. 3 in Ref.~\cite{garykarimabadi2009}, it gives a value very close to the correct value of $\gamma/|\Omega_e|\approx 0.01$ even for $k_0=1.0$, $k_{\parallel 0}=0.5$. We will estimate $\gamma(k_0,k_{\parallel 0})$ at similar values. The exponent $\alpha$ mimics the growth of the damping rate with $k_{\perp}$ at varying rates. When we take $\alpha=0$, we get the lowest bound for collisionless damping. $A(\beta_e)$ is taken to be 0.05 at $\beta_e=0.15$ from Fig.~2 of Ref.~\cite{garykarimabadi2009}. 

\begin{table}
\begin{center}
\caption{Table of various parameters and the estimates of damping rate ($|\gamma E_{>}|$) and nonlinear energy cascade rate ($|dE_{<}/dt|$)}
\label{table1}
\begin{tabular}[h]{|c|c|c|c|c|c|c|}
\hline
$t_1/\tau_A$ & $t_2/\tau_A$ & $k_0$ & $k_{\parallel 0}$ & $\alpha$ & $|dE_{<}/dt|$ & $|\gamma E_{>}|$ \\
\hline
0.4 & 0.6 & 1.0 & 0.1 & 0.0 & 1397 & 215 \\
\hline
0.4 & 0.6 & 1.0 & 0.2 & 1.0 &  1397 & 695 \\
\hline
0.6 & 0.8 & 0.8 & 0.1 & 1.5 & 943 & 389 \\
\hline
0.6 & 0.8 & 1.0 & 0.2 & 1.3 & 987 & 635 \\
\hline
0.8 & 1.0 & 1.0 & 0.1 & 1.0 & 629 & 177 \\
\hline
0.8 & 1.0 & 0.7 & 0.2 & 0.5 & 580 & 260 \\
\hline
0.8 & 1.0 & 0.8 & 0.2 & 1.5 &  600 & 532 \\
\hline 
1.0 & 1.2 & 1.0 & 0.1 & 0.0 & 325 & 75 \\
\hline
1.0 & 1.2 & 0.9 & 0.2 & 1.0 & 318 & 257 \\
\hline
1.0 & 1.2 & 0.8 & 0.2 & 1.5 & 309 & 382 \\
\hline
\end{tabular}
\end{center}
\end{table}
Table~\ref{table1} lists these estimates for different values of $t_1$, $t_2$, $k_0$, $k_{\parallel 0}$, and $\alpha$ from  the spectra of VPIC simulation run I and the linear damping estimates. The $|dE_{<}/dt|$ column gives the estimate for nonlinear energy cascade crossing $k_0$, while the $|\gamma E_{>}|$ column gives the kinetic damping estimate for $k_{\perp}>k_0$. If we choose $k_0=0.8$, $k_{\parallel}=0.2$ and $\alpha=1.5$ (the last row) we can explain all the dissipation by kinetic damping, for later times. At earlier times, the spectrum is not entirely filled in and so the kinetic damping need not balance the nonlinear transfer. We do see dissipation set in by $k_{\perp}d_i=0.8$ in Fig.~\ref{kperp_all}, so $k_0=0.8$ is a reasonable choice. We also see significant power beyond $k_{\parallel}d_i=0.2$ in our simulations, so choice of $k_{\parallel}$ is also reasonable. The damping depends strongly on $\alpha$, which is arbitrary. When $\alpha=0$, we get the lowest estimate of collisionless damping, and we see from the table that it is not sufficient to dissipate the energy being cascaded. However, the damping rate increases with $\alpha$, and we see that even at $\alpha=1.0$ the collisionless damping becomes a significant fraction of the energy cascade. Also, we have taken a single $k_{\parallel}$, but the damping rate also depends strongly on $k_{\parallel}$. From this we can say that kinetic damping has the capability to explain a large fraction of the total dissipation. To accurately calculate kinetic damping, we need the exact dependence of $\gamma$ on $k_{\perp}$ and $k_{\parallel}$ which requires a linear Vlasov solver. Nevertheless, these rough estimates show that kinetic damping can possibly be a large part of the dissipation in our VPIC simulations.  

This analysis gives us confidence that the MHD result is well-converged and the kinetic code is able to reproduce the MHD result when a larger box size is taken. It shows that the gap between fluid and kinetic descriptions of plasmas is bridged with the use of a state-of-the-art particle-in-cell code coupled with supercomputing power. The steeper spectrum seen for smaller box sizes in VPIC could be due to the  limited separation between the outer and inner scales, thus not giving enough room for the inertial range to fully develop. In Fig.~\ref{kperp_all} we see that the break in spectrum for MHD runs 0, I, and II shifts to higher wave-numbers as resolution increases, but the spectral break in corresponding VPIC runs remains fixed. This shows that the dissipative scale in ideal MHD depends on the grid resolution whereas in VPIC it is a physically meaningful length scale. Now that MHD inertial range behavior has been established in VPIC, we next investigate the behavior of thin current sheets.   

\bigskip

\noindent{\bf IV. Characterization of current sheets}

\begin{figure}[h]
\centering
\resizebox{0.5\textwidth}{!}{
\includegraphics{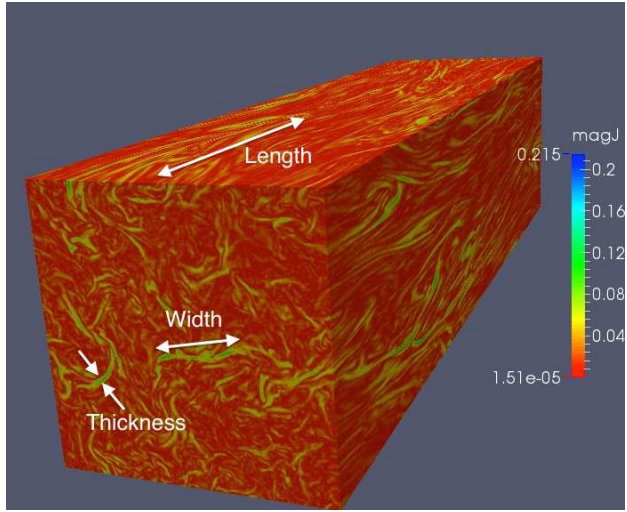}}
\caption{(color online) Surface plot of current density in VPIC run 0 at $t/\tau_A=0.8$. The three characteristic dimensions of a current sheet are illustrated. }
\label{jdensity}
\end{figure}

Several numerical studies have shown the spontaneous formation of thin current sheets in MHD turbulence. Fig.~\ref{jdensity} shows the current density on the surface of the three dimensional simulation box at $t/\tau_A=0.8$ of VPIC run 0. We see typical current sheet structures forming. There are three distinct dimensions of the sheets. First is the longest dimension along the background magnetic field, we call this the length. Then there is the intermediate dimension of width, which is the largest extent of the current sheet in the plane perpendicular to the length. The smallest dimension of the current sheet is the thickness, which is approximately perpendicular to the length and width. These three dimensions are illustrated in Fig.~\ref{jdensity}. 

\begin{figure}[h]
\centering
\resizebox{0.35\textwidth}{!}{
\includegraphics{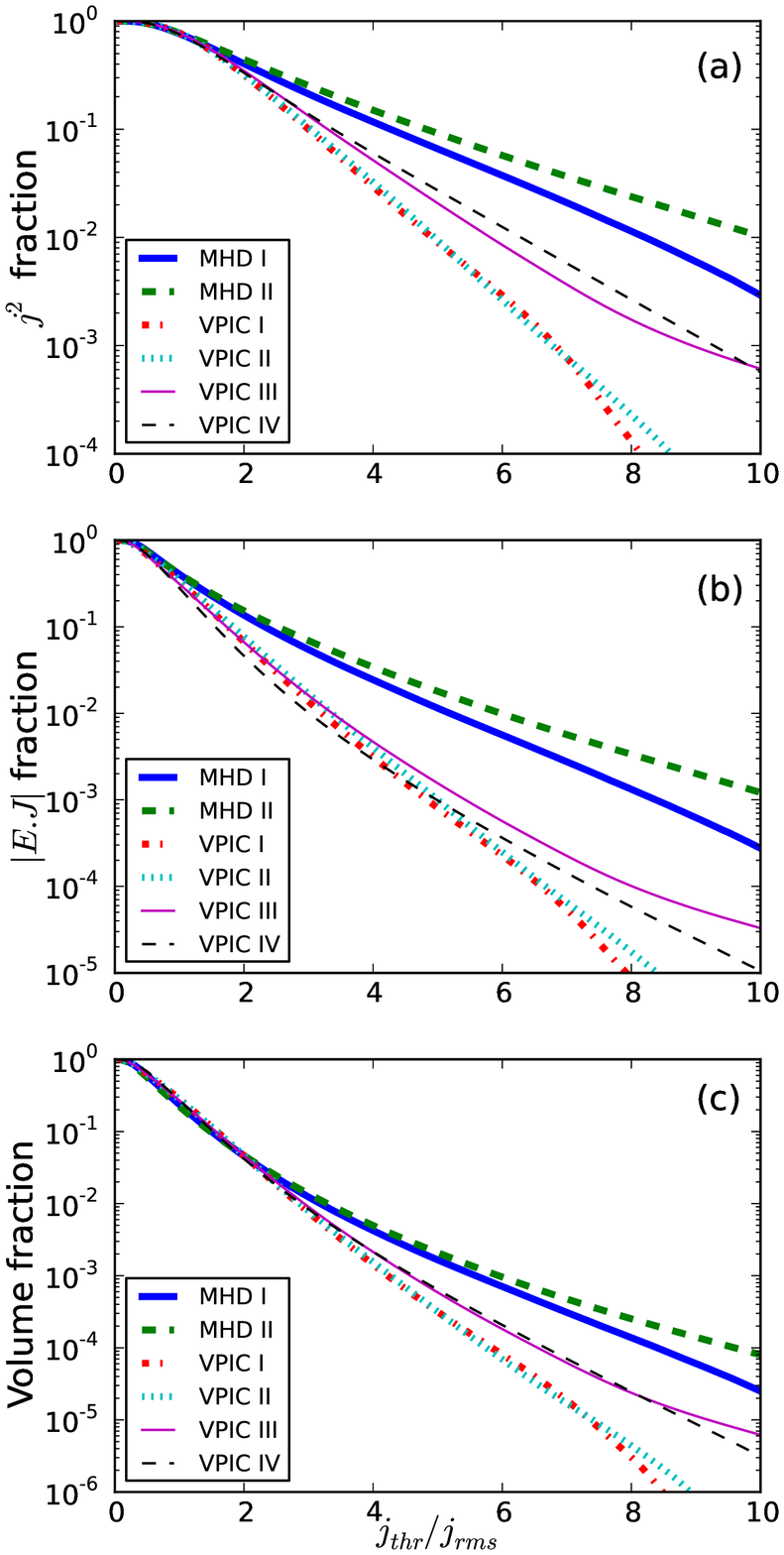}}
\caption{(color online) Fraction of (a) $|\mathbf{J}|^2$, (b) $|\mathbf{E}\cdot\mathbf{J}|$, and (c) volume, on the $y$ axis, as a function of current density above $j_{thr}$. The different curves are for the different simulations shown in the legend. These distributions are averaged for time steps between $t/\tau_A=1.0$ to $t/\tau_A=3.0$ in VPIC cases (except for VPIC case IV which goes only up to $t/\tau_A=2.6$). The MHD runs I and II are averaged at three time samples: $t/\tau_A=1.0,2.0,3.0$.}
\label{edotj_cum}
\end{figure}
Let us first look at the global distribution of current density in our simulations. In Fig.~\ref{edotj_cum}(c) we plot the volume fraction occupied by current densities above a threshold $j_{thr}$. Starting from the entire volume fraction of 1.0 for $j_{thr}/j_{rms}=0$, the distribution rapidly drops, with current density higher than $j_{rms}$ occupying as little as $20\%$ of the volume. This indicates that current density is concentrated in small structures. Current densities higher than $4j_{rms}$ occupy roughly $0.5\%$ of the volume in the MHD simulations and roughly $0.2\%$ volume in kinetic simulations. There are different proxies used to calculate energy dissipation in current density structures. In resistive MHD simulations, the energy density dissipation rate is given by $\eta j^2$. Although our simulations are not of resistive MHD, we can assume a dissipation rate proportional to $j^2$ and calculate its fraction at various current density thresholds, which is what is shown in Fig.~\ref{edotj_cum}(a). The distributions are very close to exponential till $j_{thr}/j_{rms}=8.0$. This matches results from resistive MHD simulations (see Fig. 4 in Ref.~\cite{zhdankinboldyrev2014}). In fact, the shape of the volume fraction in Fig.~\ref{edotj_cum}(c) also matches results from Ref.~\cite{zhdankinboldyrev2014}. For the MHD simulations, at $j_{thr}/j_{rms}=4.0$, we see a $j^2$ dissipation fraction of around $15\%$, and we recall that this is in a volume fraction of $0.5\%$. Similarly, VPIC simulations show a $j^2$ dissipation fraction ranging from $3\%$ to $6\%$ at the same threshold. This shows that kinetic simulations do not dissipate as strongly as MHD simulations in regions of higher current density. A similar observation is made in Ref.~\cite{tenbargehowes2013} where it is stated that due to ``\emph{the limited number of dissipation channels available in fluid simulations the energy content of small-scale current sheets is likely to be overestimated}''. This same result is borne out by Fig.~\ref{edotj_cum}(b) which plots the fraction of dissipation, now calculated using $|\mathbf{E}\cdot\mathbf{J}|$. The PIC code directly supplies us with $\mathbf{E}$, whereas for the ideal MHD code, we use $\mathbf{E}=-\mathbf{v}\times\mathbf{B}$. Above $j_{thr}/j_{rms}=4.0$, roughly $2\%$ of the $|\mathbf{E}\cdot\mathbf{J}|$ dissipation takes place in MHD codes, and $0.3\%$ of $|\mathbf{E}\cdot\mathbf{J}|$ dissipation occurs in PIC simulations. So this proxy shows even weaker concentration of dissipation in strong current sheets. One general feature common between the three Figs.~\ref{edotj_cum}(a), (b), and (c) is that changing the resolution of VPIC simulations (going from run I to run II), we do not observe any significant change in the distributions up to large $j_{thr}/j_{rms}$. On the other hand, MHD runs I and II show a difference even at low $j_{thr}/j_{rms}>2.0$. This indicates that dissipation in current sheets is sensitive to resolution in MHD but not in VPIC. On the other hand, changing the box size does affect the distribution of $j^2$ dissipation in VPIC. To understand this behavior, we take a closer look at the morphology of current sheets in our simulations.  

We can characterize the morphological properties of the current sheets using the program developed in Refs.~\cite{zhdankinuzdensky2013},~\cite{zhdankinboldyrev2014}. The algorithm of Ref.~\cite{zhdankinuzdensky2013} identifies current sheets by finding peaks in the current density and identifying the surrounding volume where the current density is higher than a certain threshold. We take the threshold to be $j_{thr}/j_{rms}=4.0$. It then finds the three major dimensions of the  sheet-like structures as shown in Fig.~\ref{jdensity}. There are two methods to characterize these dimensions, the more intuitively direct Euclidean method, and the more mathematically rigorous Minkowski method. Both these methods are described in Ref.~\cite{zhdankinboldyrev2014}. We use the Euclidean method in this work. In order to correctly identify contours of current density in VPIC, we have to smoothe the particle noise which creates irregular boundaries of the sheet structure. This is done by inline time averaging within the code. The data is averaged over an interval of $\Delta t=0.2\tau_{ci}$ during the simulation for runs 0, I, III, and IV and $\Delta t=0.1\tau_{ci}$ for run II, where $\tau_{ci}=2\pi/\omega_{ci}$. This averaging procedure significantly reduces the particle noise, giving smooth boundaries of current sheets. This data has been used for all the analysis in this work. For MHD the current density is calculated by taking $\nabla\times \mathbf{B}$. In VPIC we can also obtain the current density directly from particle velocity. Both the methods agree very well in VPIC, and we use $\nabla\times\mathbf{B}$ to calculate current density in both codes, for consistency.  

\begin{figure}[h]
\centering
\resizebox{0.5\textwidth}{!}{
\includegraphics{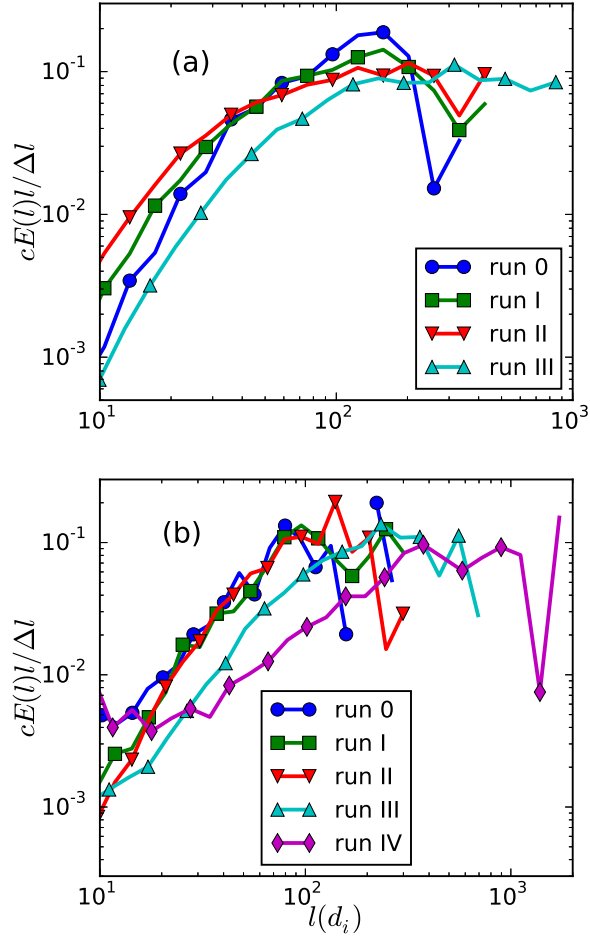}}
\caption{(color online) Length distributions in the different simulation runs. The x-axis is normalized to $d_i$. (a) is for MHD runs, (b) is for VPIC runs. }
\label{length_plot}
\end{figure}

Fig.~\ref{length_plot} plots an intensity weighted distribution function of the lengths (the longest dimension) of current sheets. The distribution function is $cE(l)l/\Delta l$ plotted versus $l$, where $E(l)$ is the total intensity in current sheets of length $l$. The intensity is the $j^2$ dissipation rate in the current sheet,  defined as $\int_{l} d^3x j^2$, where the integral is over the volume of current sheets with length between $l-(\Delta l/2)$ and $l+(\Delta l/2)$. The intensity $E(l)$ is multiplied by $l$ because this makes the maximum correspond to the length scale where most of the overall $j^2$
dissipation is organized. The constant $c$ is defined to normalize  the distribution function to unity, $\sum_{l}(cE(l)l/\Delta l)=1$. The width of the bins $\Delta l$ is proportional to length $l$ in order to make uniform bins on a log scale. This distribution is averaged over the 11 time snap shots from $t=1.0\tau_A$ to $t=3.0\tau_A$ for each simulation (except for run IV of VPIC, in which the average runs only up to $t=2.6\tau_A$). 

In Fig.~\ref{length_plot}(a) we see the length distributions for  MHD runs 0, I, II, and III. The maximum intensity occurs in current sheets with lengths starting from $100d_i$ and going up to the box length in $z$, with the distribution almost flat between these two values. Runs 0, I and II show very similar distributions, showing that the lengths do not depend on the resolution. Run III shows a length distribution which looks very similar to run II, except it is shifted to the right by a factor of two. This reflects the fact that run III in MHD is the same as run II, only difference being that it's length normalization is a factor of two larger than the normalization of run II. These results are seen more clearly in Fig.~\ref{length_plot}(b) which shows the length distributions for VPIC runs. We see that the distributions for runs 0, I, and II are very similar to each other, even at small length scales. The distribution reaches a peak at around $100d_i$ and extends up to $l\approx 300 d_i$, not going up to the entire box length in $z$. Below $l\approx 20d_i$, we see the distributions flatten out for runs 0 and IV. This is due to the PIC noise, as we see this flattening most prominently for these two runs which have the lowest resolution, and least prominently for run II which has the highest resolution. As we double the box size in run III, we find the distribution shifting to the right with the most energetic sheets having a length between $200d_i$ and $600d_i$. Similarly, doubling the box size further in run IV, the sheets go up to length $1800d_i$. However, the distribution drops off at $l\sim 1200d_i$, which is double of the maximum length observed in Run III of $600d_i$. This is consistent with the doubling of maximum length from $300d_i$ to $600d_i$ in going from Run II to Run III. The spike at $1800d_i$ is due to percolating structures, i.e., structures that extend throughout the simulation domain, which would require a higher threshold to get a cleaner scaling. It should be noted that skin-depth is a natural normalization in kinetic codes, and hence changing box size is not just a change in normalization. The driving wavenumbers of turbulence defined in Eqs.~\ref{deltab} \& \ref{deltav} are defined with respect to the box length, and so changing box size implies changing the driving scale of turbulence. The results from both codes clearly show that the lengths of current sheets scale linearly with the driving scale of turbulence. It is expected that there should be a limit on the maximum length of current sheets, beyond which they would break-up due to some instability, like the current sheet sausage instability~\cite{leewang1988} or kink instability~\cite{daughton1998}. However, it appears that neither our ideal MHD nor VPIC simulations have reached such a limit as our distributions can be seen doubling and quadrupling with the box length.

\begin{figure}[h]
\centering
\resizebox{0.5\textwidth}{!}{
\includegraphics{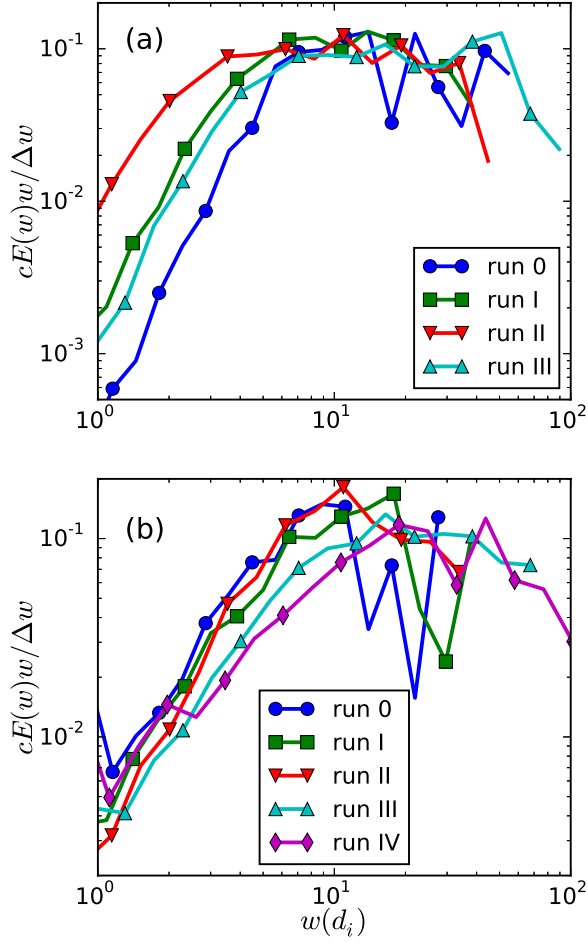}}
\caption{(color online) Distributions of current sheet widths for different simulation runs. (a) is for MHD and (b) is for VPIC.}
\label{width_plot}
\end{figure}
Fig.~\ref{width_plot} shows the intensity-weighted width distribution for the different runs, defined as $cE(w)w/\Delta w$, analogous to the length distribution, with $w$ representing width. MHD runs 0, I, and II show maximum intensity in the range from $w\approx 7d_i$ to $w\approx 40d_i$. The distribution falls off steeply below this range. The fall-off depends on the resolution, with the higher resolution runs showing the fall-off at smaller widths. The distributions do not extend to $w=170 d_i$ which is the maximum possible width in the transverse direction (considering the diagonal in the transverse direction which is $120\sqrt{2}d_i$). Doubling the box size in run III just shifts the distribution given by run II by a factor of two, just as was seen for the length distribution in Fig.~\ref{length_plot}(a). The width distributions for VPIC runs in Fig.~\ref{width_plot}(b) agree well with MHD. Runs 0, I, and II show more peaked distributions compared to MHD, with the most intense sheets having a width around $10d_i$. The distribution falls off around this value, implying a well-defined range for current sheet widths. Unlike in MHD, the fall-off at lower widths does not vary with the resolution. Compared to these three runs, run III with double box length shows widths that are larger but not exactly double. It  also has a broader distribution of the most intense widths ranging from $w\approx 20d_i$ to $w\approx 70 d_i$. This broader distribution is similar to the distribution in MHD run III. Going back to Fig.~\ref{length_plot}, we observe that the MHD and VPIC length distributions also match better for run III compared to runs 0, I, and II. We recall that the inertial range spectra between MHD and VPIC also agreed closely for runs III and IV in Fig.~\ref{kperp_all}. We now see that the current sheet distributions also agree closely for these runs, giving more confidence that the VPIC simulations are reaching the MHD limit. When we double the box length again in run IV, we again see a slight shift of the distribution to the right, but it is not by a factor of two. Although the distribution extends up to $w=100d_i$, it falls off above $w\approx 70d_i$. Thus, the widths appear to scale with the driving scale, but the scaling is weaker than linear. One possibility is that the current sheets might be breaking up due to tearing instability, limiting their width. This will need further investigation.   

\begin{figure}[h]
\centering
\resizebox{0.5\textwidth}{!}{
\includegraphics{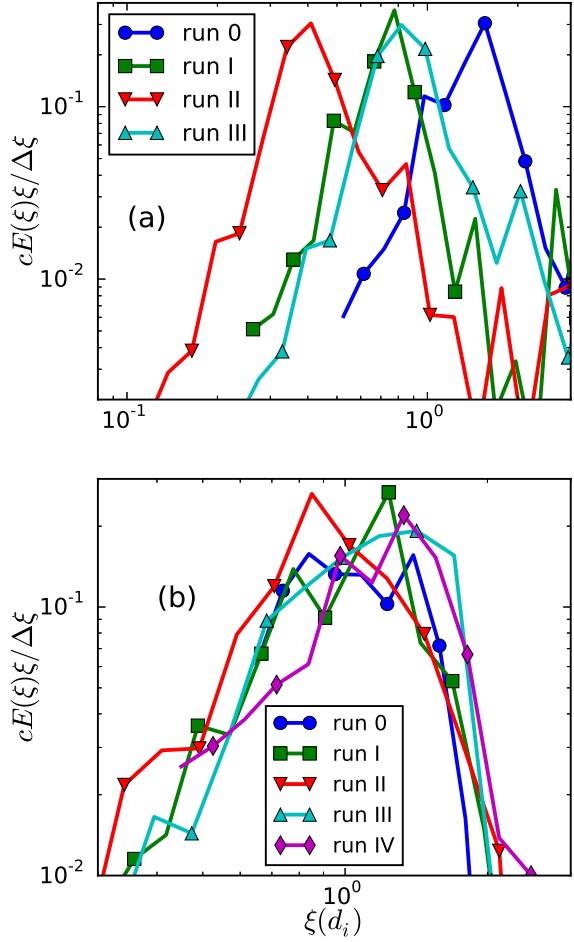}}
\caption{(color online) Distributions of current sheet thickness for different simulation runs. (a) is for MHD and (b) is for VPIC.}
\label{thickness_plot}
\end{figure}
Fig.~\ref{thickness_plot} plots the intensity weighted thicknesses ($\xi$, the smallest dimension) for all the simulations. Thickness shows very distinct features in both the codes. The distributions are extremely peaked in both the codes, implying an extremely well-defined thickness for current sheets. MHD run 0 shows a peak thickness of around $1.5d_i$. Run I shows a peak thickness of around $0.8d_i$ which is half of the run 0 value. This shows that current sheet thickness in our MHD simulations is entirely dependent on the grid spacing. The grid resolution determines the numerical diffusivity in our runs and that sets the thickness, so the current sheets become as thin as the grid allows them to be. This is further seen in run II where the resolution is again doubled and the thickness reduces further by half to become $0.4d_i$. Run III has the same resolution as run I. Thus, it shows the same current sheet thickness as run I. In resistive MHD simulations analyzed in Ref.~\cite{zhdankinboldyrev2014} the dissipation is controlled by specifying an explicit resistivity. In those cases it is found that current sheet thickness depends on the resistivity, smaller the resistivity, thinner the sheets. Thus, sheet thickness depends on the dissipation; if it is explicit then it depends on the diffusion coefficient, if implicit then it depends on the grid spacing.  In VPIC, the sheets thickness is remarkably converged in all the cases. Apart from the statistical fluctuations, the distributions show a peak close to $1d_i$. This is irrespective of change in resolution (runs 0, I, and II) or change in box size (runs II, III, and IV). This indicates that VPIC is simulating the physical mechanism limiting the current sheet thickness, and our resolution is enough to capture it. Moreover, the most intense current sheets show a thickness of $1d_i$. This proves that skin-depth is the limiting factor determining current sheet thickness. 


\bigskip

\noindent{\bf V. Summary and Discussion}

Numerical simulations of MHD turbulence are extremely valuable in understanding several laboratory, space, and astrophysical systems. To understand dissipation of MHD turbulence better simulation capabilities are needed. At scales smaller than the  ion gyro-radius or ion skin-depth, we expect the MHD framework to breakdown, and hence use of kinetic codes becomes imperative. This work has attempted to bridge the gap between fluid and kinetic simulations of MHD turbulence. We found that it is possible to generate MHD turbulence using a fully kinetic code, which reproduces several features of fluid simulations. 

The energy dynamics are remarkably similar in both codes despite the vast difference in the way they handle dissipation - PLUTO was run in ideal MHD mode with only numerical diffusivity, whereas VPIC is fully kinetic. This tells us that the idea that small scale turbulence adjusts to dissipate whatever the rate of energy cascade down from the inertial range is holds. In this paper, we have started our simulations purely with shear-Alfven waves. There is some conversion into compressible modes as seen from the formation of density fluctuations. It would be interesting to see how well the PIC and fluid energy dynamics match for other initial conditions. The rate of energy decay is very close to $E\sim t^{-1}$. This rate of decay is also observed in other MHD simulations of decaying turbulence with zero magnetic helicity~\cite{biskampmuller1999}, which is the case for our ensemble of shear-Alfven waves. It would be interesting to see if VPIC is able to reproduce the decay rate of $E\sim t^{-0.5}$ for simulations with finite magnetic helicity, and even simulations with high $\beta$. 

Power spectra have been analyzed for incompressible MHD turbulence in a large number of studies~\cite{boldyrevperez2011},~\cite{beresnyak2014}. In this work, we have investigated spectra from compressible MHD and particle-in-cell codes. All the simulations show well-converged spectra under a change of resolution. The decaying turbulence presents its own challenges in analyzing spectra, but they are managed reasonably by averaging with a time-dependent compensating factor. The MHD simulations show a spectral slope of $k_{\perp}^{-1.3}$, for all the simulation runs. The smaller box runs of VPIC show a steeper spectra of $k_{\perp}^{-1.5}$, but they are converged w.r.t. resolution. The slope decreases and arrives at the MHD value of $k_{\perp}^{-1.3}$ for larger box size in VPIC. This shows that provided sufficient separation of scales between the driving scale and dissipative scale, VPIC is able to reproduce the MHD inertial range. Typically, MHD turbulent spectra are observed to be between $k_{\perp}^{-3/2}$ and $k_{\perp}^{-5/3}$. Our spectra show shallower slope, with reasons not clear. The decaying nature of the turbulence plus the compressibility can affect the spectrum. Although our initial condition contains only shear-Alfven waves which are incompressible, we observe generation of density fluctuations later in time. This has been observed in other simulations also, in similar parameter regimes, where Alfven waves interact nonlinearly to produce the compressible fast and slow modes~\cite{stoneostriker1998},~\cite{cholazarian2002}. Preliminary simulations of forced turbulence show the $k_{\perp}^{-1.5}$ slope. The VPIC and MHD spectra differ at wave numbers $k_{\perp}d_i\gtrsim1.0$. The magnetic fluctuations are strongly damped in VPIC compared to MHD. Making use of collisionless damping rates from the linear Vlasov equation, we estimated the rate of energy dissipation by kinetic damping. It seems plausible that kinetic damping is dissipating a significant fraction of the energy in our PIC simulations.  

Current sheets have been proposed as the dominant dissipative structures in MHD turbulence~\cite{servidiomatthaeus2009}. Techniques have been developed to identify such structures and analyze them~\cite{zhdankinboldyrev2014}. We use such techniques to identify and characterize current sheets. We find sheets which are vastly elongated in the direction parallel to the mean field. Their lengths range from $100 d_i$ extending up to the entire length of the simulation box. The fluid and kinetic codes match well for the length distributions. The lengths scale linearly with the driving scale of turbulence in both codes. The widths show a similar behavior as the length scales. The most intense sheets have widths ranging from $10d_i$ upwards, roughly a tenth of their lengths. They show a scaling with the outer driving scales of the turbulence, however it is weaker than a linear scaling. It is expected that sheets with a large aspect ratio of width to thickness would be unstable to tearing modes, and this might play in role in setting the widths of current sheets. Current sheet thickness shows distinctly different behavior in both the codes. It depends entirely on the grid spacing in our ideal MHD simulations, in which dissipation is entirely provided by the finite grid resolution. It does not scale with the driving scale of turbulence. In VPIC, it is very close to the skin-depth, depending neither on resolution nor on the driving scale. This implies a fundamental, kinetic mechanism that limits the thickness of current sheets to skin-depth.

Overall, we have found that with large box sizes and sufficient number of particles, particle-in-cell code VPIC remarkably reproduces a wide variety of MHD turbulent dynamics in the inertial range. Conversely, this work also confirms that MHD describes real turbulence at large scales, consistent with kinetic simulations using first principles. The important differences due to kinetic physics begin to show up at the current sheet thickness scale of skin-depth. So VPIC is able to capture the entire range of dynamics from inertial to dissipative range. This opens up the possibility of studying the problem in a comprehensive manner. We can look at the large scale dynamics governing current sheets and also relate it to the microphysics going on at kinetic scales. This will be helpful in understanding conversion of magnetic and kinetic energy into heat in MHD turbulence. Heating in our simulations will be studied in detail in a future work.    

\bigskip

\noindent{\bf Appendix}

\begin{figure}[h]
\centering
\resizebox{0.6\textwidth}{!}{
\includegraphics{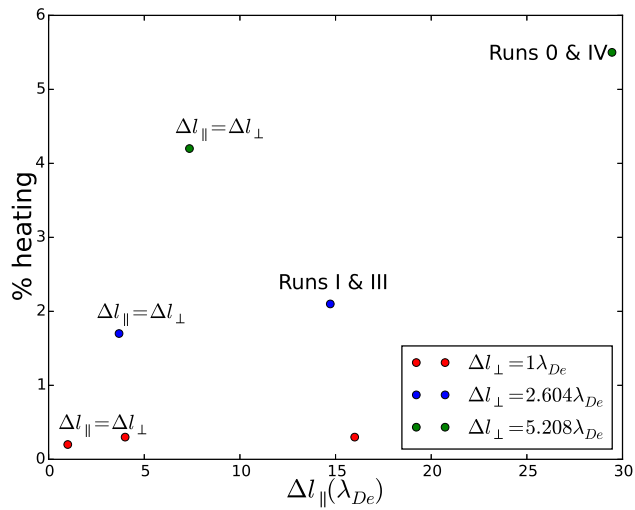}}
\caption{The parallel cell size ($\Delta l_{\parallel}$) is varied for different perpendicular cell sizes ($\Delta l_{\perp}$). The different colors show different $\Delta l_{\perp}$ while the x-axis plots different $\Delta l_{\parallel}$. The y-axis show the heating as a percent of the initial thermal energy. Points with the same resolution as our simulation runs 0, I, III, and IV are marked accordingly.}
\label{debye1}
\end{figure}

The parallel cell size in all our simulations and the perpendicular cell size in runs 0 and IV is going above the $\Delta x=3.0\lambda_{De}$ limit for the finite grid instability. This is calculated using the relation $\lambda_{De}=(v_{th}/c)d_e=0.08d_e$, where $\lambda_{De}$ is the electron Debye length. The finite grid instability should exhibit itself as numerical heating of the plasma. To find out how important the finite-grid instability is in our simulations we do a test of the numerical heating by taking a small box with $20\times 20\times 20$ cells in each direction, introducing the same plasma parameters, but without the turbulent fluctuations. We let the simulation run for the same amount of time as run I (in terms of $\tau_{pe}$). We then observe how much the plasma heats up. We begin with a cell size of 1 electron Debye length ($\lambda_{De}$) in all three directions. Then we degrade the resolution by increasing the cell size in both directions till we reach our simulation resolution. The percentage of numerical heating as a fraction of the  initial internal energy is shown in Fig.~\ref{debye1}.  
The blue dots have the same perpendicular resolution as runs I and III. Green dots have same perpendicular resolution as runs 0 and IV. Runs I and III show very little heating (2.2\%). Run II has even lesser heating as its resolution is higher. Runs 0 and IV are our worst resolved simulations and they show a heating of $5.5\%$. This shows that there is some finite-grid instability led heating going on for our runs, but it is very weak and should not affect our results in any significant way. It should be pointed out that the finite-grid instability for PIC codes has not been studied extensively for pair plasmas and also for different cell sizes in parallel and perpendicular directions. Our simulations seem to suggest that for pair plasmas the resolution in parallel direction can be relaxed above $\Delta l_{\parallel}=3.0\lambda_{De}$. 

\noindent{\bf Acknowledgments}

This work was supported in part by the National Science Foundation sponsored Center for Magnetic Self Organization (CMSO) at the University of Chicago. HL gratefully acknowledges the support by the LANL/LDRD program and the DOE/Office of Fusion Energy Science through CMSO. The simulations were performed on the Bluewaters supercomputer at the National Center for Supercomputer Applications at the University of Illinois at Urbana-Champaign.

\newpage

\end{document}